# Surface Bloch waves mediated heat transfer between two photonic crystals

Philippe Ben-Abdallah[1], Karl Joulain[2] and Andrey Pryamikov[2]

1) Laboratoire de Thermocinétique, CNRS UMR 6607, Ecole Polytechnique de l'Université de Nantes, 44 306 Nantes cedex 03, France.

2) Institut P', CNRS-Université de Poitiers-CNRS UPR 3346, 86022 Poitiers Cedex, France.

**Abstract**

We theoretically investigate the non-radiative heat transfer between two photonic crystals separated by a small gap in non-equilibrium thermal situation. We predict that the surface Bloch states coupling supported by these media can make heat exchanges larger than those measured at the same separation distance between two massive homogeneous materials made with the elementary components of photonic crystals. These results could find broad applications in near-field technologies.

[1] Electronic mail : pba@univ-nantes.fr

[2] Electronic mail : karl.joulain@univ-poitiers.fr

Classical radiometry theory[1] predicts that the net flux exchanged between two hot bodies cannot exceed the value exchanged between two perfect blackbodies. At subwavelength distances, the situation radically changes and this theory fails to describe heat transfers[2,3] due to the presence of non-radiative (non-propagative) electromagnetic modes which coexist with the radiative (propagative) ones. In these conditions, evanescent modes localized close to the surface become the main contributors to energy transfer so that it becomes possible to exceed the far field limit [4,5]. In particular, when two massive materials support resonantly coupled surface modes such as surface polaritons, heat transfer is able to surpass by several orders of magnitude the limit predicted by the Planck theory. This singular behaviour has opened possibilities for the development of innovative near-field technologies such as near-field thermophotovoltaïc conversion[6], plasmon assisted nanophotolitography[7] or near-field spectroscopy[8].

Recently this subwavelength heat transfer theory has also been applied to massive materials coated by thin films[9-10] to be finally extended to arbitrary layered materials[11]. In the present Letter, we apply this theoretical framework to investigate near-field heat exchanges between some specific multilayered media, the periodic layered materials, also called photonic crystals (PCs). We show that near-field heat exchanges between two PCs in non-equibrium thermal situation are completely driven by the surface Bloch waves coupling and can be controlled by an appropriate design of PCs.

Let us first consider the system described in Fig.1 which is composed of two arbitrary multilayered materials of total thickness $e_L = \sum_{i=1}^{M-1} d_i$ and $e_R = \sum_{i=M+1}^{N} d_i$ deposed on two transparent (i.e. non-emitting) massive materials. These coated materials are separated by a vacuum gap of thickness $d_M$. We also suppose, for the seek of simplicity, that all involved materials are nonmagnetic and have a frequency dependent complex dielectric functions of the form $\varepsilon_k(\omega) = \varepsilon'_k(\omega) - i\varepsilon''_k(\omega)$ with $\varepsilon''_k > 0$ for $k = 1,...,N$. In addition, we assume that the left and right media are maintained in nonequilibrium thermal situation with two fictive thermal baths of temperatures $T_L$ and $T_R$, respectively. Due to the presence of thermal fluctuations within these media, the local charges inside any layer k randomly oscillate to give rise, in other layers j, to electric and magnetic fields $\mathbf{E}_{k\to j}$ and $\mathbf{H}_{k\to j}$. The corresponding energy flux in layer j is given by the Poynting vector

$$\mathbf{S}_{k\to j}(\mathbf{r},\omega) \equiv 2\,\text{Re}\{\mathbf{E}_{k\to j}(\mathbf{r},\omega)\times \mathbf{H}^{*}_{k\to j}(\mathbf{r},\omega)\}. \quad (1)$$

Since, the energy absorbed by a body is equal to the total flux which crosses its boundary, the net radiative flux exchanged between the left and right coating reads in terms of total Poynting vector $\mathbf{S}_{j}(\mathbf{r},\omega) = \sum_{k}\mathbf{S}_{k\to j}(\mathbf{r},\omega)$ (the summation operates over all random sources which effectively contribute to the energy balance)

$$\varphi_{LR}(\omega) = \{\langle \mathbf{S}_{M+1}((e_{L}+d_{M})^{+},\omega)\rangle - \langle \mathbf{S}_{N}((e_{L}+e_{R}+d_{M})^{-},\omega)\rangle - \langle \mathbf{S}_{M-1}(e_{L}^{-},\omega)\rangle + \langle \mathbf{S}_{1}(0^{+},\omega)\rangle\}\mathbf{e}_{z}. \quad (2)$$

In this expression $\langle\cdot\rangle$ denotes the statistical averaging over all realizations of random sources and $\mathbf{e}_{z}$ is the unit vector normal to each layer. To calculate each term of this relation, we proceed as followed. Due to the linearity of Maxwell equations[12,13], the total electric field $\mathbf{E}_{j} = \sum_{k}\mathbf{E}_{k\to j}$ and magnetic field $\mathbf{H}_{j} = \sum_{k}\mathbf{H}_{k\to j}$ in layer j can be related to the local fluctuating currents $\mathbf{j}_{k}(\mathbf{r}_{s},\omega)$ by the following expressions

$$\mathbf{E}_{j}(\mathbf{r},\omega) = -i\mu_{0}\omega \sum_{k}\int_{V}d\mathbf{r}_{s}\,\overline{\overline{\mathbf{G}}}_{E}^{jk}(\mathbf{r},\mathbf{r}_{s},\omega)\cdot\mathbf{j}_{k}(\mathbf{r}_{s},\omega), \quad (3\text{-a})$$

$$\mathbf{H}_{j}(\mathbf{r},\omega) = \sum_{k}\int_{V}d\mathbf{r}_{s}\,\overline{\overline{\mathbf{G}}}_{H}^{jk}(\mathbf{r},\mathbf{r}_{s},\omega)\cdot\mathbf{j}_{k}(\mathbf{r}_{s},\omega), \quad (3\text{-b})$$

where $\overline{\overline{\mathbf{G}}}_{E}^{jk}(\mathbf{r},\mathbf{r}_{s},\omega)$ and $\overline{\overline{\mathbf{G}}}_{H}^{jk}(\mathbf{r},\mathbf{r}_{s},\omega)$ denote the electric and magnetic dyadic Green tensors between the source layer and the observation layer. It follows from these expressions that, the i[th] component of total Poynting vector inside layer j writes

$$\langle\mathbf{S}_{j}(\mathbf{r},\omega)\rangle\cdot\mathbf{e}_{i} = -2\,\text{Re}\left\{i\omega\mu_{0}\eta_{ilm}\sum_{k}\int_{V}\int_{V}\overline{\overline{\mathbf{G}}}_{El\alpha}^{jk}(\mathbf{r},\mathbf{r}',\omega)\overline{\overline{\mathbf{G}}}_{Hm\beta}^{jk\,*}(\mathbf{r},\mathbf{r}'',\omega)\langle j_{k\alpha}(\mathbf{r}',\omega)j_{k\beta}^{*}(\mathbf{r}'',\omega)\rangle d\mathbf{r}'d\mathbf{r}''\right\} \quad (4)$$

where the curl operator has been expressed in term of Levi-Civitas tensor $\eta$. Here we have adopted the Einstein summation convention over the repeated index $\alpha$, $\beta$, $l$ and $m$ which denote the Green tensors components.

According to the fluctuation dissipation theorem[14], the fluctuating current cross correlation function reads

$$\langle j_{k\alpha}(\mathbf{r},\omega) \bullet j_{k\beta}^*(\mathbf{r}',\omega) \rangle = \frac{\omega \varepsilon_0 \varepsilon_k''(\omega) \Theta(\omega, T_{L,R})}{\pi} \delta_{\alpha\beta} \delta(\mathbf{r} - \mathbf{r}'),  \qquad (5)$$

with $\Theta(\omega, T_{L,R}) \equiv \hbar\omega/[\exp(\hbar\omega/k_B T_{L,R}) - 1]$ the mean energy of a Planck oscillator at the equilibrium temperature $T_L$ or $T_R$ depending on the considered coating. Then, using this expression and the expansion of Green tensors in terms of the intracavity fields[11] the net flux exchanged between both media can be recast into the simple form

$$\varphi_{LR}(\omega) = \frac{1}{\pi^2} [\Theta(T_L, \omega) - \Theta(T_R, \omega)]$$

$$\times \sum_{\lambda=s,p} \left\{ \underbrace{\frac{1}{4} \int_{k_{//}<\frac{\omega}{c}} dk_{//} \frac{(1-|r_L^\lambda|^2)(1-|r_R^\lambda|^2)}{|1 - r_L^\lambda r_R^\lambda e^{-2i\beta_M' d_M}|^2} k_{//}}_{\text{propagative}} + \underbrace{\int_{k_{//}>\frac{\omega}{c}} dk_{//} \frac{\operatorname{Im}(r_L^\lambda) \operatorname{Im}(r_R^\lambda)}{|1 - r_L^\lambda r_R^\lambda e^{2\beta_M'' d_M}|^2} k_{//} e^{2\beta_M'' d_M}}_{\text{evanescent}} \right\} \qquad (6)$$

where $r_L^\lambda$ and $r_R^\lambda$ are the reflexion coefficients the cavity sides and $\beta_M = \beta_M' + i\beta_M''$ is the normal component of wavevector in this layer (by convention $\beta_M'' < 0$). The first integral denotes the contribution of propagative modes (i.e. $k_{//} < \omega/c$) while the second one is that of non-propagative ones.

Let us now pay our attention on the non-propagative exchanges between two periodic layered media made with high contrast index materials, the so called photonic crystals[15]. To investigate qualitatively and quantitatively these exchanges, it is convenient to introduce the monochromatic non-radiative heat transfer coefficient

$$h(\omega; T_R) \equiv \lim_{T_L \to T_R} \left| \frac{\varphi_{LR}(\omega)}{T_L - T_R} \right|_{k_{//}>\omega/c} \qquad (7)$$

defined as the ratio of net radiative flux exchanged over the coating temperature difference when this discrepancy tends to zero. For concreteness, we first consider here two quarter-wave coatings made with alternating layers of transparent material and lossy dielectric with thickness $d_1 = 1\mu m$ and $d_2 = 1.66 \mu m$ and dielectric constants $\varepsilon_1 = 6.25$ and $\varepsilon_2 = 2.25 - 0.3i$, respectively. We also assume that the number of period is finite but can be arbitrary large. In infinite periodic structures, we know that Bloch waves are eigenstates. The dispersion relations for these modes for both polarizations are given by[16]

$$\cos(K_z \Lambda) = \cos(\beta_1 d_1)\cos(\beta_2 d_2) - \xi_q \sin(\beta_1 d_1)\sin(\beta_2 d_2) \tag{8}$$

where

$$\xi_q = \begin{cases} \frac{1}{2}(\frac{\beta_1}{\beta_2} + \frac{\beta_2}{\beta_1}) & \text{, in polarization s} \\ \frac{1}{2}(\frac{\varepsilon_2 \beta_1}{\varepsilon_1 \beta_2} + \frac{\varepsilon_1 \beta_2}{\varepsilon_2 \beta_1}) & \text{, in polarization p} \end{cases}$$

and $\Lambda = d_1 + d_2$ is the period of the structure. At any frequency $\omega$, a real solution for $K_z$ represents a propagating Bloch state. On the other hand, when $|\cos(K_z \Lambda)| > 1$, the Bloch states are evanescent waves. These waves belong to the so-called forbidden bands of the periodic medium. The collection of all Bloch states in the $\omega - k_{//}$ plane defines the band structure of material. In Fig. 2-a, band structure of Bloch modes in a quarter-wave photonic crystal is displayed in the mid-infrared region around the Wien's frequency $\omega_W = 2.821 k_B T / \hbar$ at 300K. The straight line represents the light line in vacuum and the grey bands frame the allowed zones. In a finite photonic crystal all confined Bloch modes are discrete modes. The local density of state (LDOS) calculated from the reflection coefficient of crystal[17] (Fig. 2-b and 2-c) shows that these modes are evanescent in the surrounding vacuum and their number increases with the period number of PC.

Heat transfer coefficients between these quarter waves PCs and several other coupled systems made with identical PCs are plotted in Fig. 3-a at different frequencies. They show that the transfer is maximal in the spectral region where the surface Bloch waves have been highlighted. Moreover, we note in Fig. 3-b, that the heat transfer coefficient exhibit a maximum at a certain distance contrary with what is traditionally observed (Fig. 3-c) between two massive materials. The origin of this behaviour still remains an open question. However, one can speculate that it is the result of complex interactions between the different Bloch modes throughout the vacuum gap. This suggests that the heat transfer might be optimized by an appropriate design of PCs. On the other hand and more interestingly, we see in Figs. 3-b,c that the magnitude of heat transfer can surpass the values observed between two non structured massive materials at the same separation distance. As confirmed in Fig. 3-a, this original behaviour is due to the presence of several surface modes which couple and create channels for heat exchanges between both materials. As the number of mode increases, energy exchange also is magnified. Finally, due to the confinement of surface Bloch waves close to the light line ($\beta^{//} << \omega/c$), we see in Figs. 3-b,c that the enhancement of transfer beyond the blackbody limit can be observed at longer separation distances than between two massive materials. Moreover, this behaviour has been observed simultaneously for both polarization states [only s-polarisation is plotted in Figs. 3].

In summary, we have seen that the near-field heat transfers between PCs are driven by a surface Bloch waves coupling. In comparison with massive materials, we have shown that these confined modes allow magnifying heat exchanges by a factor of three in both polarization states. In addition, transfer enhancement takes place at separations distances ten times larger than usually. Consequently, PCs-based coatings seem promising structures to improve the performance of numerous near-field technologies such as the near-field thermophotovoltaïc energy conversion or the near-field thermophotolitography.

**Captions List**

Fig.1 Structure consisting of plane layers made of absorbing materials coating deposited on semi-infinite transparent materials. These textured materials are separated by a vacuum gap and are maintained in nonequilibrium thermal situation at temperatures $T_L$ and $T_R$, respectively.

Fig.2 (a) Band structure of a quarter-wave infinite photonic crystal ($\varepsilon_1 = 6.25$, $\varepsilon_2 = 2.25 - 0.3i$ and $d_1 = 1\mu m$, $d_2 = 1,66\mu m$) for TE waves. White zones are forbidden bands in which $|\cos(K_z \Lambda)| > 1$. The black straight line delimits the light cone in vacuum. (b) LDOS of evanescent modes at $z = 50 nm$ from the surface plotted in the $\omega - k_{//}$ plane for a quarter-wave photonic crystal with 5 periods and 15 periods (c). The white triangular zone which corresponds to propagative modes is not detailed.

Fig.3 (a) Monochromatic non-radiative heat transfer coefficient for TE waves (at 300 K) between two photonic crystals (case I) made with 15 periods of unit cell with $\varepsilon_1 = 16$, $\varepsilon_2 = 2.25 - 0.3i$ $d_1 = 1,1\mu m$ and $d_2 = 0,55\mu m$ and (case II) between two quarter-wave photonic crystals made with 5 and 15 periods ($\varepsilon_1 = 6.25$, $\varepsilon_2 = 2.25 - 0.3i$) respectively and for a 100 nm separation distance. The dashed curve describes exchanges between two massive materials of dielectric constant $\varepsilon_2 = 2.25 - 0.3i$ at the same separation distance. Heat transfer coefficient in case I (b) and between two massive materials (c) vs the separation distance.

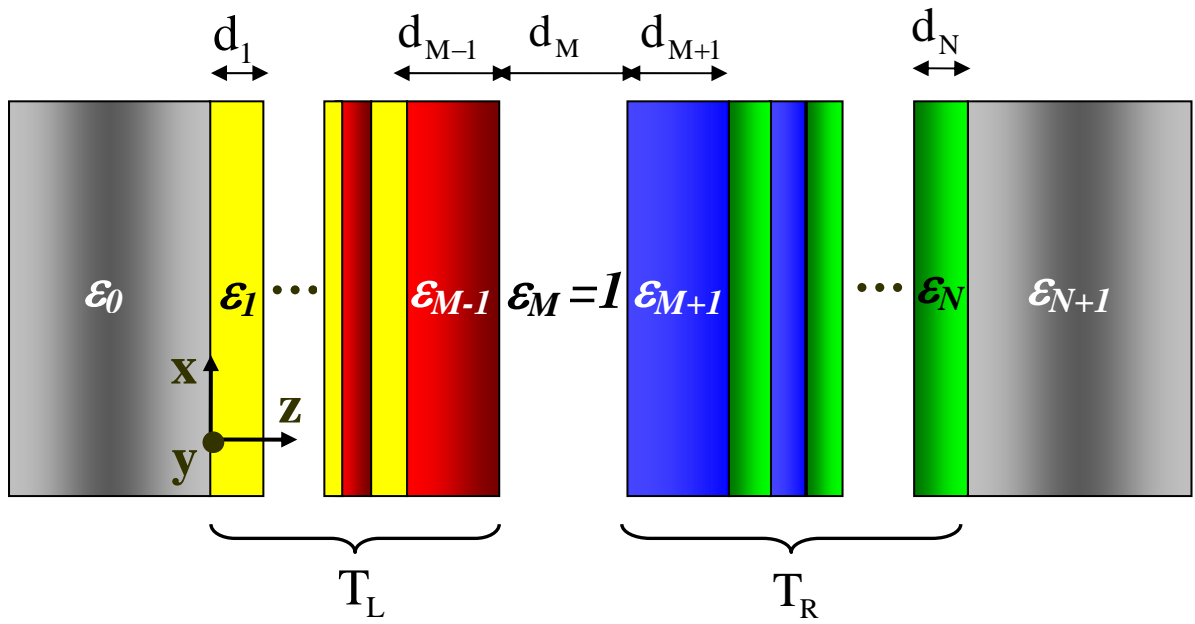

**Figure 1**

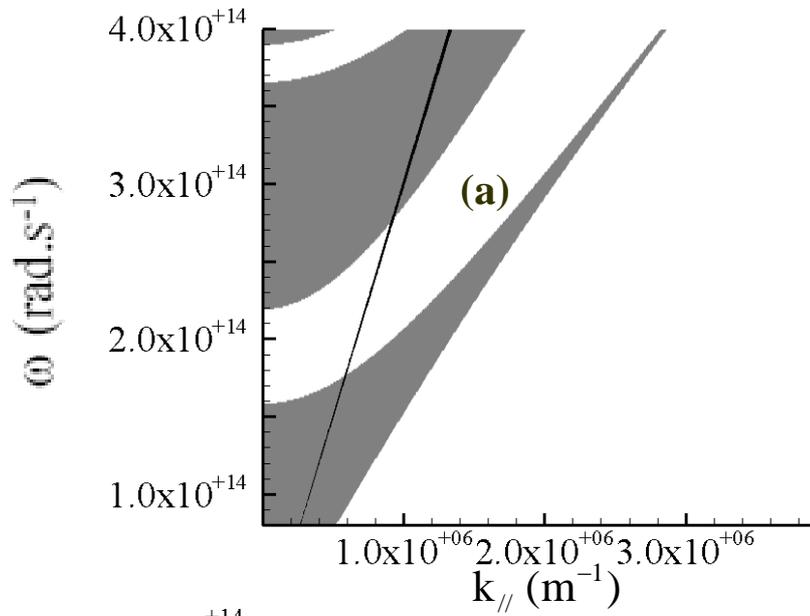
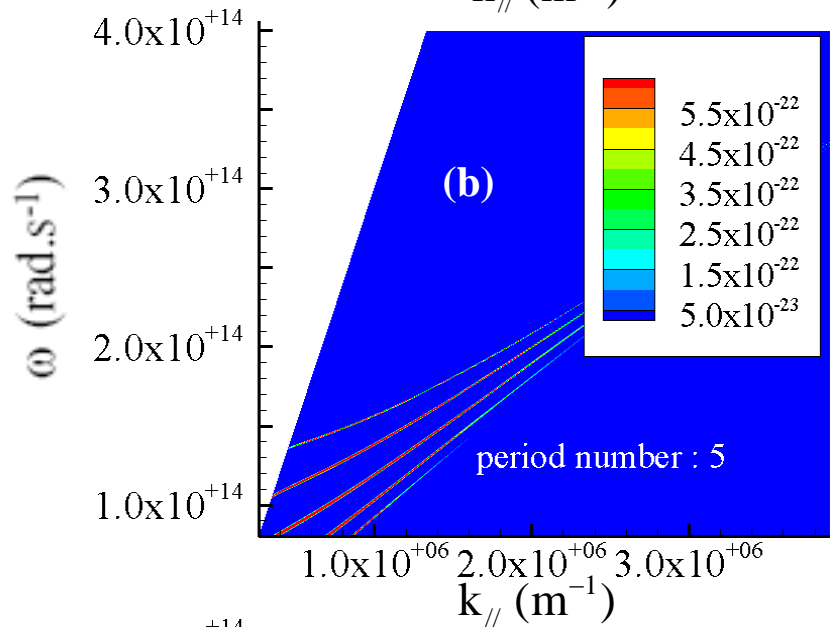
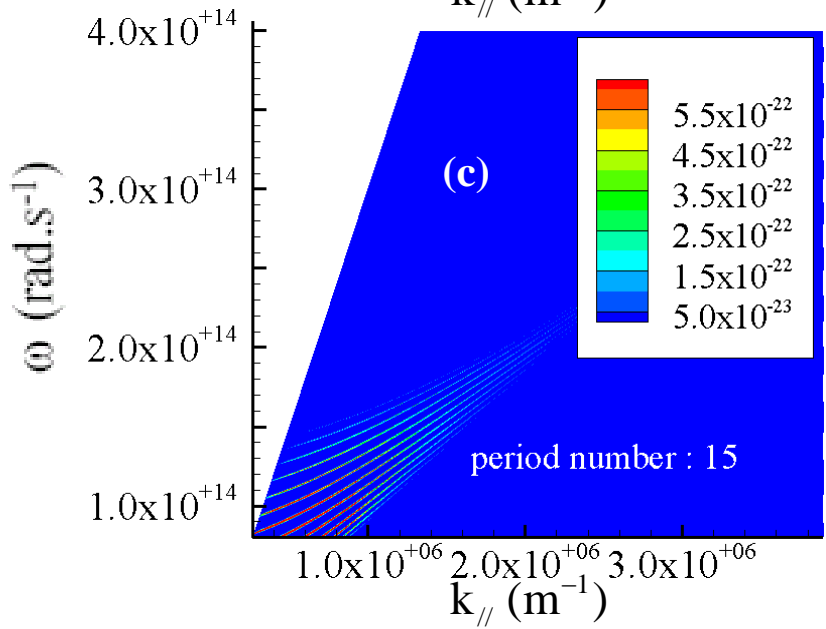

**Figure 2**

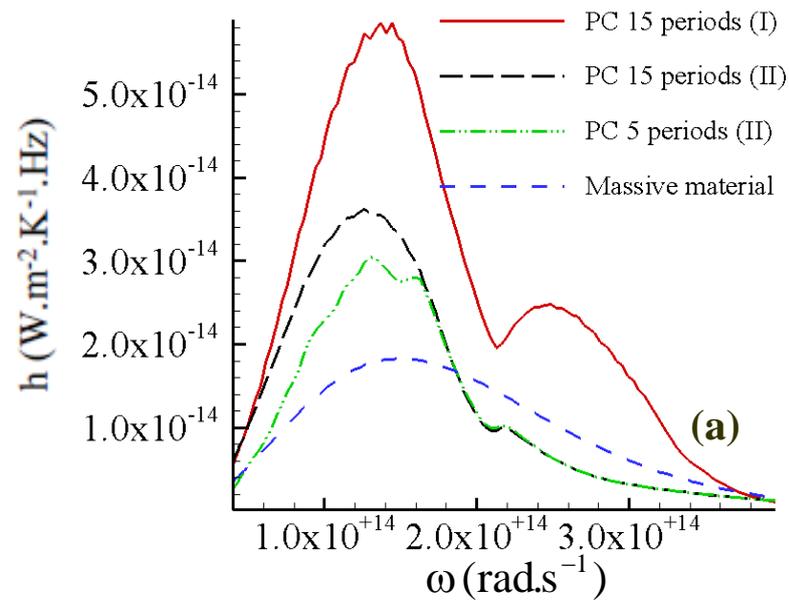

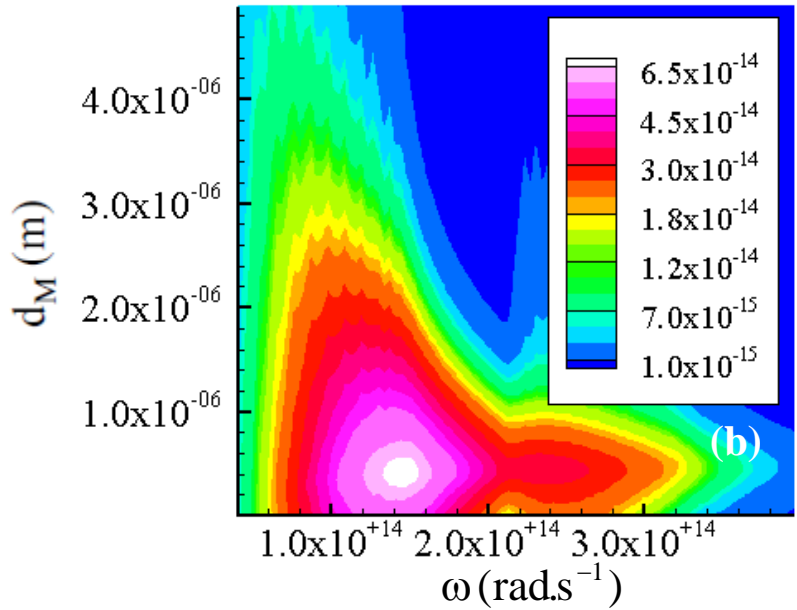

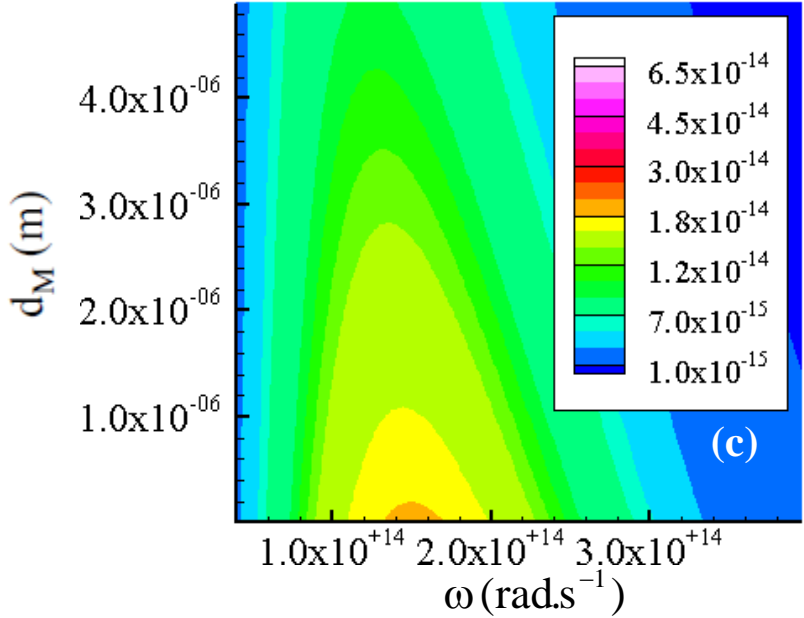

Figure 3